\documentclass[twocolumn,tighten]{aastex63}

\usepackage{savesym}
\savesymbol{tablenum}
\usepackage{siunitx}
\restoresymbol{SIX}{tablenum}

\newcommand{\halpha}{H$\alpha$}

\newcommand{\hei}{\ion{He}{1}}

\newcommand{\henir}{{\ion{He}{1}}$~\lambda$10830}

\newcommand{\mdot}{$\dot{\text{M}}$}
\newcommand{\Mdot}{{\rm \dot{{M}}}}

\newcommand{\msunyr}{\rm{M_{\sun} \, yr^{-1}}}

\newcommand{\kms}{\rm \, km \, s^{-1}}
\newcommand{\ri}{R$_{\rm i}$}
\newcommand{\rw}{W$_{\rm r}$}
\newcommand{\tmax}{T$_{\rm max}$}

\def\curf{{\cal F}}

\newcommand{\newarcsec}[1]{\ang[angle-symbol-over-decimal]{;;#1}}

\received{---- --, ----}
\revised{---- --, ----}
\accepted{---- --, ----}
\submitjournal{ApJ}

\shorttitle{Variable Accretion of PDS 70}
\shortauthors{Thanathibodee et al.}

\begin{document}

\title{Variable Accretion onto Protoplanet Host Star PDS 70}

\correspondingauthor{Thanawuth Thanathibodee}
\email{thanathi@umich.edu}

\author[0000-0003-4507-1710]{Thanawuth Thanathibodee}
\affiliation{Department of Astronomy, University of Michigan, 1085 South University Avenue, Ann Arbor, MI 48109, USA}

\author{Brandon Molina}
\affiliation{Department of Climate and Space Sciences and Engineering, University of Michigan, 2455 Hayward Street, Ann Arbor, MI 48109, USA}

\author[0000-0002-3950-5386]{Nuria Calvet}
\affiliation{Department of Astronomy, University of Michigan, 1085 South University Avenue, Ann Arbor, MI 48109, USA}

\author[0000-0001-7351-6540]{Javier Serna}
\affiliation{Instituto de Astronom\'ia, Universidad Nacional Aut\'onoma de M\'exico, Ensenada, M\'exico}

\author[0000-0001-7258-770X]{Jaehan Bae}
\affiliation{Department of Terrestrial Magnetism, Carnegie Institution for Science, 5241 Broad Branch Road NW, Washington, DC 20015, USA}

\author[0000-0003-1621-9392]{Mark Reynolds}
\affiliation{Department of Astronomy, University of Michigan, 1085 South University Avenue, Ann Arbor, MI 48109, USA}

\author[0000-0001-9797-5661]{Jes\'us Hern\'andez}
\affiliation{Instituto de Astronom\'ia, Universidad Nacional Aut\'onoma de M\'exico, Ensenada, M\'exico}

\author[0000-0002-5943-1222]{James Muzerolle}
\affiliation{Space Telescope Science Institute, 3700 San Martin Drive, Baltimore, MD 21218, USA}

\author[0000-0002-1650-3740]{Ramiro Franco Hern\'andez}
\affiliation{Instituto de Astronom\'ia y Meteorolog\'ia, Universidad de Guadalajara, Avenida Vallarta No. 2602, Col. Arcos Vallarta, CP 44130, Guadalajara, Jalisco, Mexico}

\begin{abstract}
The PDS 70 system has been subject to many studies in the past year following the discovery of two accreting planets in the gap of its circumstellar disk. Nevertheless, the mass accretion rate onto the star is still not well known. Here we determined the stellar mass accretion rate and its variability based on TESS and HARPS observations. The stellar light curve shows a strong signal with a $3.03\pm0.06$ days period, which we attribute to stellar rotation.  Our analysis of the HARPS spectra shows a rotational velocity of $v\sin\,i=16.0\pm0.5\,{\rm km\,s^{-1}}$, indicating that the inclination of the rotation axis is $50\pm8$ degrees.  This implies that the rotation axes of the star and its circumstellar disk are parallel within the measurement error. We apply magnetospheric accretion models to fit the profiles of the H$\alpha$ line and derive mass accretion rates onto the star in the range of $0.6-2.2\times10^{-10}\,{\rm M_{\odot}yr^{-1}}$, varying over the rotation phase. The measured accretion rates are in agreement with those estimated from NUV fluxes using accretion shock models. The derived accretion rates are higher than expected from the disk mass and planets properties for the low values of the viscous parameter $\alpha$ suggested by recent studies, potentially pointing to an additional mass reservoir in the inner disk to feed the accretion, such as a dead zone.  We find that the He\,I\,$\lambda$10830 line shows a blueshifted absorption feature, indicative of a wind. The mass-loss rate estimated from the line depth is consistent with an accretion-driven inner disk MHD wind.
\end{abstract}

\keywords{Accretion, H I line emission, Protoplanetary disks, T Tauri stars}

\section{Introduction} \label{sec:intro}
The pre-main sequence star PDS~70 has attracted 
much recent attention
because it hosts the most unambiguous example of planets in the process of formation, namely two giant planets inside the gap of the circumstellar disk
\citep{keppler2018,haffert2019}. 
Indicators that the planets are still forming include the submillimeter emission detected around them, interpreted as arising in circumplanetary disks \citep[CPDs,][]{isella2019},
and by the {\halpha} emission coincident with the location of the protoplanets in NIR images 
\citep{wagner2018,haffert2019}, interpreted as forming in accretion flows from the circumstellar disks into the planet-CPD systems 
\citep{thanathibodee2019b,aoyama2019}.

PDS~70 is a K7 star \citep{riaud2006,pecaut2016} located in the 5-10 Myr old Upper Sco association \citep{gregorio-hetem2002}. Its circumstellar disk has been classified as a pre-transitional disk \citep{espaillat2008},  based on the large size ($\sim$ 65\,au) of the cavity in the disk 
inferred from the spectral energy distribution (SED) of the system and consistent with NIR imaging
\citep{dong2012}. 
\citet{gregorio-hetem2002} classified the star as a Weak T Tauri star (WTTS) based on an {\halpha} equivalent width of 2 {\AA}, and thus identified it as a non-accretor \citep{white2003,barrado-y-navascues2003}. More recently, 
\citet{long2018a} classified the star as a non accretor, based on the lack of emission in Pa$\beta$. Similarly, \citet{joyce2019} found essentially no excess over
the photosphere in the Swift U band flux of PDS~70, from which they infer that the star is accreting at a extremely low level, 
$\Mdot \sim 6\times10^{-12} \msunyr$, consistent with no accretion.
On the other hand, \citet{haffert2019} reported
a redshifted absorption component in the {\halpha} profile of the star and concluded that the star is accreting.

\citet{thanathibodee2018,thanathibodee2019a} are carrying out a program searching for the lowest accretors; preliminary results of this program find that around
20-30 \% of stars previously classified as WTTS are still accreting, presumably at very low levels
(Thanathibodee et al., in prep.). The new accretors in this program have been identified using the 
{\henir} line.
Redshifted absorption components at velocities consistent with free-fall, which are formed in 
magnetospheric accretion flows, may appear in this line even at low densities, due to the metastable nature of its lower level. Therefore, the {\henir} line can detect very low levels of accretion, even in cases when traditional accretion tracers, such as {\halpha}, exhibit no sign of stellar accretion.
\citep{thanathibodee2019a}.

Given the different assessments in the literature, in this paper, we revisit the accretion status of PDS~70. In agreement with \citet{haffert2019}, we find that it is accreting mass from the disk. We also obtain a reliable estimate of the mass accretion rate and its variability. The mass accretion onto the star is important because it is a key constraint on models of planet formation. For instance, \citet{zhu2011} found that multiple planets were needed to create cavities of the order of tens of au, but in this case, mass coming from the outer disk would be partitioned between the planets and little would reach the star; however, this was at odds with the relatively high mass accretion rates onto the star measured in transitional disks \citep[cf.][]{espaillat2014}. The constraints imposed by the mass accretion rate can be particularly useful for a case so well characterized as PDS~70. In this case, we have a well studied circumstellar disk \citep{dong2012,keppler2019}, we have clear detection of planets with approximate mass and radius \citep{keppler2018,muller2018}, and we also have estimates of the mass of the circumplanetary disks \citep{isella2019} and the mass accretion rates onto the planets \citep{thanathibodee2019b,haffert2019,aoyama2019}. With a good estimate of the mass accretion rate onto the star, the system will be ideal for testing models of planet formation and planet-disk interaction.

This paper is organized as follows. In \S 2 we describe the observations and data sources. In \S 3 we derive stellar properties and apply magnetospheric models to fit {\halpha} profiles and derive the mass accretion rate onto the star and its variability. In \S 4, we discuss the implications of our results. Finally, in \S 5 we give our conclusions.

\section{Observations and Data Sources}

\begin{deluxetable}{llccc}[t]
\tablecaption{Summary of Observations and Data Sources \label{tab:obs}}
\tablehead{
\colhead{Instrument} 	&
\colhead{Obs. Date} &
\colhead{Exp. time} &
\colhead{No. of} &
\colhead{SNR\tablenotemark{a}} \\
\colhead{} 		&
\colhead{(UT)} 	&
\colhead{(sec)} &
\colhead{Obs.} &
\colhead{}
}
\startdata
FIRE 	& 2019 Apr 26   & 253.6  & 1     & 260 \\
TESS    & 2019 Apr 26 to	& $2 \times 60$    & 13,887  & 570 \\
        & 2019 May 20 	&                   &   &  \\
HARPS   & 2018 Mar 29  &   900  & 6  & 15 \\
        & 2018 Mar 30  &   900  & 6  & 11 \\
        & 2018 Mar 31  &   900  & 6 & 18 \\
        & 2018 Apr 18  &  1800  & 1 & 21 \\
        & 2018 Apr 19  &   900  & 2 &  8 \\
        & 2018 Apr 20  &  1800  & 2 & 22 \\
        & 2018 Apr 21  &  1800  & 1 & 22 \\
        & 2018 Apr 22  &  1800  & 1 & 22 \\
        & 2018 Apr 23  &  1800  & 1 & 22 \\
        & 2018 May 01  &  1800  & 2 & 14 \\
        & 2018 May 06  &  1800  & 2 & 26 \\
        & 2018 May 13  &  1800  & 2 & 17 \\
\enddata
\tablenotetext{a}{Signal-to-noise at 10830\,{\AA} for FIRE, median SNR for HARPS.}
\end{deluxetable}

\subsection{NIR Spectroscopy}
We obtained a near-infrared spectrum of PDS~70 using the FIRE spectrograph on the Magellan Baade Telescope at the Las Campanas Observatory in Chile. With the \newarcsec{0.6} slit, the spectrograph achieved a resolution of $\sim$6000 for simultaneous spectral coverage in the range 0.9~-~2.4$\mu$m. We observed the star in two nodding positions (A/B), each with an exposure time of 126.8\,s in the Sample-Up-The-Ramp (SUTR) mode. A telluric standard star was observed immediately afterward. The average airmass and seeing of the observation were 1.1 and \newarcsec{1.0}, respectively.  The spectral extraction, wavelength calibration, and telluric correction were performed using the IDL-based FIRE data reduction pipeline \citep{simcoe2013}.

\subsection{TESS}
\begin{figure*}[t]
\epsscale{1.1}
\plotone{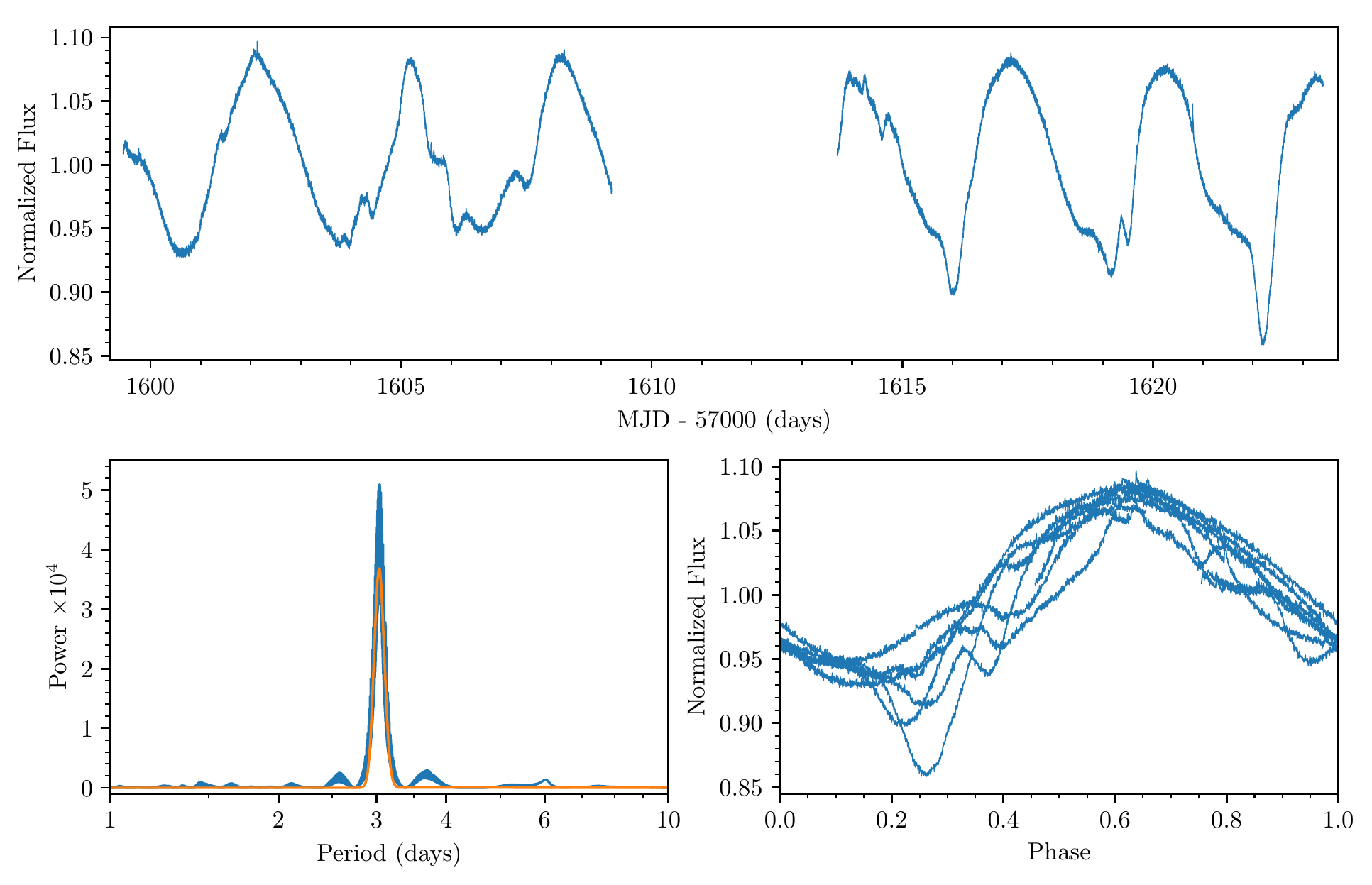}
\caption{Top: TESS light curve of PDS~70. The flux has been normalized to the median flux. Bottom left: Lomb-Scargle periodogram of the light curve (blue). The periodogram shows a strong peak at 3.03\,d. The orange line shows the Gaussian fit to the power spectrum. Bottom right: The phase-folded light curve using the period $P=3.03$\,d.
\label{fig:tess}}
\end{figure*}

PDS~70 was observed with TESS (TIC179413040) with 2 minutes cadence for 24 days, and the photometry was reduced by the TESS pipeline. We downloaded the extracted light curve from the Mikulski Archive for Space Telescopes (MAST). The Simple Aperture Photometry (SAP) flux and the Pre-search Data Conditioning SAP flux (PDCSAP) from the pipeline appear to be the same for the star, and we choose to use the PDCSAP flux. We use the {\tt Lightkurve} software \citep{lightkurve-collaboration2018} to extract and normalize the light curve. Figure~\ref{fig:tess} shows the 24-day light curve of PDS~70.

\subsection{Optical Spectroscopy}
We downloaded spectra of PDS~70, obtained with the HARPS instrument \citep{mayor2003} on the European Southern Observatory (ESO) 3.6m telescope (Program ID 098.C-0739, PI: Lagrange), from the ESO Archive Science Portal.
HARPS provided 32 high-resolution (R$\sim$115,000) spectra of the star observed across $\sim$2 months in 2018. We used the data automatically reduced and calibrated by the HARPS pipeline for our analysis.
Table 1 shows the details of the observations.

\section{Analysis \& Results}
\subsection{Stellar Properties}
\subsubsection{Rotation Period} \label{sssec:rotation_period}
To determine the rotation period of PDS~70, we constructed a periodogram from the TESS light curve, using the {\tt Psearch} code \citep{saha2017}, which combines the Lomb-Scargle and Lafler–Kinman methods. As shown in the bottom left panel of Figure~\ref{fig:tess}, the periodogram shows a strong power at the period of $3.03$ days. To estimate the uncertainty of the period, we fit a Gaussian to the strongest peak in the periodogram \citep{venuti2017}. The corresponding period from the fit is $3.03\pm0.06$ days, where the uncertainty is the $1\sigma$ propagation of the fitted Gaussian width in frequency space.

We visually inspect the period by folding the light curve with the calculated period of 3.03\,d. As shown in the bottom right panel of Figure~\ref{fig:tess}, the period is consistent with the light curve. The amplitude of the light curve is on the order of 10\%, consistent with variation due to starspots phasing in and out of the line of sight \citep{herbst1994,venuti2017}. Smaller dips can be seen at phase $\sim0.3-0.4$, suggesting that there could be other factors that modulate the light curve, such as obscuration of the inner disk warp \citep{bouvier2003}. The 3.03\,d period is also consistent with a typical rotation period of T Tauri stars \citep[e.g.][]{karim2016}. Therefore we conclude that the period of the light curve is the rotation period of the star since other types of variations would modulate the light curve in different timescales \citep[e.g.][]{siwak2018}.

We calculate the corotation radius, at which the Keplerian orbital period is equal to the stellar rotation period, and outside of which accretion cannot occur \citep{hartmann2016}. Using the stellar mass ${\rm M_{\star} = 0.76\pm0.02\,M_{\odot}}$ and radius ${\rm R_{\star} = 1.26\pm0.15\,R_{\odot}}$ \citep{muller2018}, we obtain the corotation radius as
\begin{equation}
    R_{co} = \left(\frac{GM_{\star}P^2}{4\pi^2}\right)^{1/3} = 6.4\pm0.8\,R_{\star},
\end{equation}
where the uncertainty is calculated with the standard error propagation.

\subsubsection{Rotational Velocity} \label{sssec:rotation_velocity}
We used the Fourier method \citep{carroll1933} to calculate the projected rotational velocity ($v\sin i$) of the star. The method requires isolated photospheric lines at sufficient SNR at a high spectral resolution to have a reliable line shape \citep{simon-diaz2007}. Since none of the HARPS spectra has enough SNR for the analysis, co-adding the spectra is required. We first calculated the radial velocity (RV) by cross-correlating each spectrum with a PHOENIX photospheric template \citep{husser2013} and fitting a Gaussian to the cross-correlation function. The median and the standard deviation of the RV is $+6.0\pm1.5\,\kms$. We corrected all 32 spectra with RV=$6.0\,\kms$, and combined them by averaging the spectra weighted by the median SNR (c.f. Table~\ref{tab:obs}). The resulting spectrum has a median SNR of $\sim46$. We then selected 20 photospheric lines that are clearly isolated, including the \ion{Li}{1}\,$\lambda6708$ line, and calculated a Fourier power spectrum for each line. From the first zero in the Fourier spectrum, we calculated the $v\sin i$, adopting a limb darkening coefficient $\epsilon=0.6$, which is appropriate for spectra in the optical range \citep{claret2000}. Based on the median and the standard deviation of $v\sin i$ measured from these 20 lines, the rotational velocity of the star is $v\sin(i) = 16.0 \pm 0.5\,\kms$. 

We combined our measurement of $ v \, {\rm sin}i$
with the rotation period $P$ from Section~\ref{sssec:rotation_period} and R$_{\star}=1.26\pm0.15$\,R$_{\odot}$ \citep{muller2018}, 
to obtain the inclination of the star as
\begin{equation}
    i = \sin^{-1}\left(\frac{P
\, v\sin(i)}{2\pi R_{\star}}\right) = 50\pm8^{\circ}~(1\sigma).
\end{equation}
\citet{keppler2019} found 
$i=51.7\pm0.1^{\circ}$ and $i=52.1\pm0.1^{\circ}$
for the inclination of the protoplanetary disk in two different ALMA observations. Therefore,  
our results suggest that the stellar and disk rotation axes are parallel to each other within the measurement errors.

\subsection{Accretion status of PDS~70 from FIRE observations}
Optical hydrogen lines such as {\halpha} have been used to determine if a T Tauri star is still accreting. However, {\halpha}
may fail at very low accretion rates due to a comparatively significant contribution from chromospheric emission \citep{manara2017}, especially when observed in low or moderate spectral resolution.
On the other hand, the {\henir} line
has been found to be very sensitive to accretion and thus to be a good tracer for low-accretors \citep{thanathibodee2018}. In particular, the presence of a redshifted absorption component in the line profile is a direct indicator of accretion. Here we use this line to probe the state of accretion of PDS~70. 

\begin{figure*}[t]
\epsscale{1.15}
\plotone{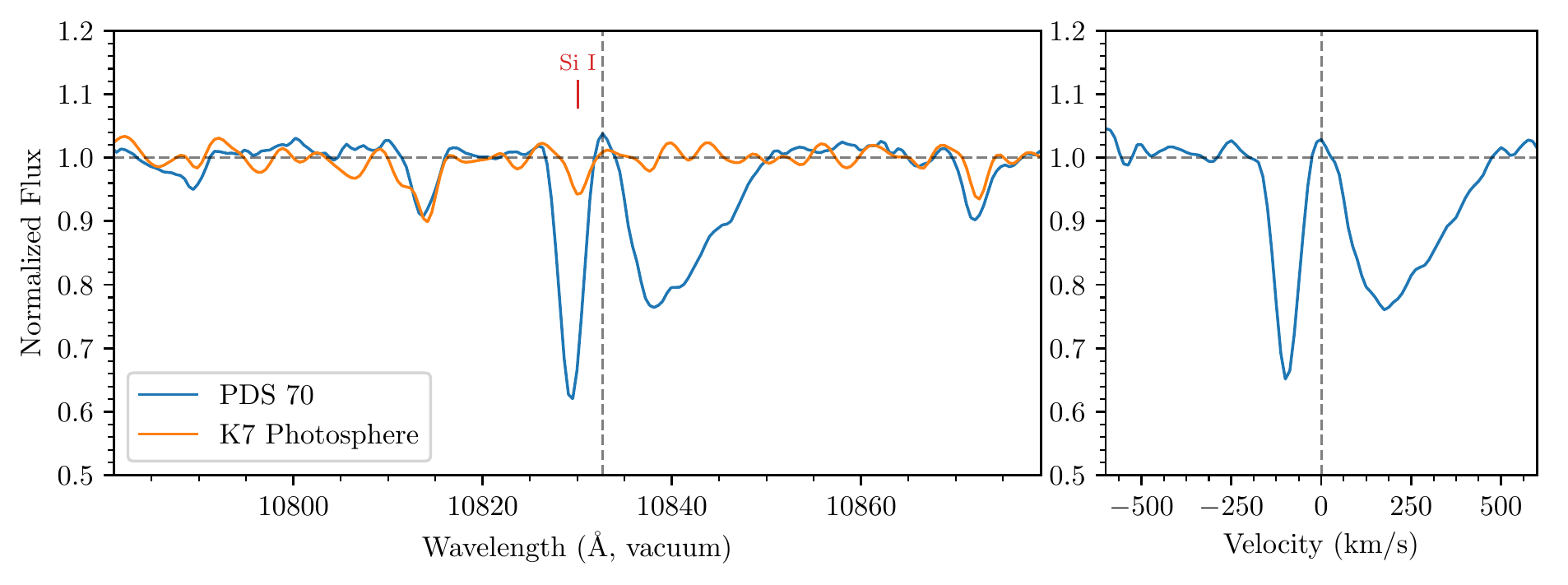}
\caption{{\henir} line profile of PDS~70. Left: The line before subtraction of the photosphere. The \ion{Si}{1} line is indicated. The nominal line center, calculated by averaging the frequency of the {\henir} triplet weighted by the $gf$ values, is shown as the vertical dashed line. Right: The line after photospheric subtraction. The line shows both blueshifted and redshifted absorption components.
\label{fig:hei}}
\end{figure*}

The {\henir} line is shown in Figure~\ref{fig:hei}.
\footnote{The analysis of the FIRE data is performed in the vacuum wavelength, and the vacuum line center of the {\henir} is at 10833\,\AA. Nevertheless, we will refer to the line using the air wavelength following the standard convention.} 
In general, the {\henir} line is conspicuous in accreting stars. However, for low accretors observed at a medium or low spectral resolution, the \ion{Si}{1} at 10830\,\AA (vacuum) could contaminate the {\hei} line at 10833\,\AA. Therefore, we constructed a photospheric template of the star from interpolating in a PHOENIX model spectrum \citep{husser2013} with the same effective temperature and gravity as PDS~70, and convolved it first to the rotational velocity of the star and then to the resolution of the FIRE spectrograph. As shown in the left panel of Figure~\ref{fig:hei}, the contribution from the \ion{Si}{1} to the overall absorption of the {\hei} line is negligible. The right panel of Figure~\ref{fig:hei} shows the line after subtracting the photospheric template. The line shows strong and conspicuous redshifted and blueshifted absorption features. The presence of the redshifted absorption feature is a definitive indication that the star is accreting. As with many low accretors with this type of profile, the {\hei} emission is weak or undetectable (Thanathibodee et al., in prep).

\subsection{Measurement of the Mass Accretion Rate}
At very low levels of accretion,
the most reliable way to measure the mass accretion rate 
is to model the resolved profile of emission lines \citep{thanathibodee2019a}. High resolution 
is needed to distinguish the chromospheric feature of the line, which 
appears as a narrow 
and mostly symmetric feature at the line center, and magnetospheric features that extend out to the star's free-fall velocity. Here we use the magnetospheric accretion model of \citet{muzerolle2001} to model the {\halpha} line profiles from the HARPS spectra.

\subsubsection{Magnetospheric Accretion Model}

\begin{deluxetable}{lccc}[t]
\tablecaption{Range of Model Parameters \label{tab:model_param}}
\tablehead{
\colhead{Parameters} & \colhead{Min.} & \colhead{Max.} & \colhead{Step} 
}
\startdata
{\mdot} ($10^{-10}\msunyr$)	& 0.2	& 4.5 	& 0.1, 0.5  \\
T$_{\rm max}$ (K)	        & 10000	& 12000	& 250 \\
R$_{\rm i}$	(R$_{\star}$)	& 2.0	& 6.0	& 0.4 \\
W$_{\rm r}$ (R$_{\star}$)	& 0.2   & 2.0   & 0.4 \\
$i$					        & 30	& 75	& 5 \\ 
\hline
\enddata
\end{deluxetable}

The physics 
of the magnetospheric accretion models is 
given in detail in \citet{hartmann1994} and
\citet{muzerolle1998a,muzerolle2001}. Here we 
describe the basic assumptions.
The model assumes that the accreting material flows along 
the magnetic field of the star,
taken as dipolar. It assumes that the stellar rotation axis, the dipolar magnetic axis, and the disk Keplerian rotation axis are aligned. As a result, the accretion flow is axisymmetric. 
The free parameters of the model are the disk truncation radius (\ri), the radial width of the accretion flow on the equatorial plane (\rw), and the maximum temperature of the gas in the flow (\tmax). 
The density in the flow is set by the mass accretion rate, {\mdot}, and the geometry.
We solve a 16-level hydrogen atom, in which the mean intensities
for the radiative rates are calculated with 
the extended Sobolev approximation, to obtain level populations and source functions.
The line profile is calculated using 
a ray-by-ray method for a given inclination ($i$) between the magnetic axis and the line of sight.

We calculate a large grid of models varying the model parameters, as shown in Table~\ref{tab:model_param}. Although we have a measurement of the inclination of the rotation axis (see Section \ref{sssec:rotation_velocity}), we probe a larger range of inclinations to verify the axisymmetric assumption of the model; differences between the model inclination and the rotational inclination would suggest a possible misalignment between the rotation axis and the magnetic axis. We only calculate models in which the outer radius of the flow, {\ri+\rw}, is inside the corotation radius of the star ${\rm R_{co}=6.4\,R_{\star}}$. The range of {\mdot} is selected to cover a typical range at which {\halpha} starts to fail as an accretion diagnostics (few $\times10^{-10}\,\msunyr$;  \citealt{thanathibodee2018}). The lower end of the {\mdot} range is chosen from our pre-grid calculations, in which the {\halpha} profiles can still be in emission. The range of the flow temperatures is selected to be consistent with the expected {\tmax} at low {\mdot} \citep{muzerolle2001,thanathibodee2019a}. In total, we calculate 64,800 models.

\subsubsection{Fitting H$\alpha$ Line Profiles} \label{sssec:fit_halpha}

\begin{figure*}[t]
\epsscale{1.15}
\plotone{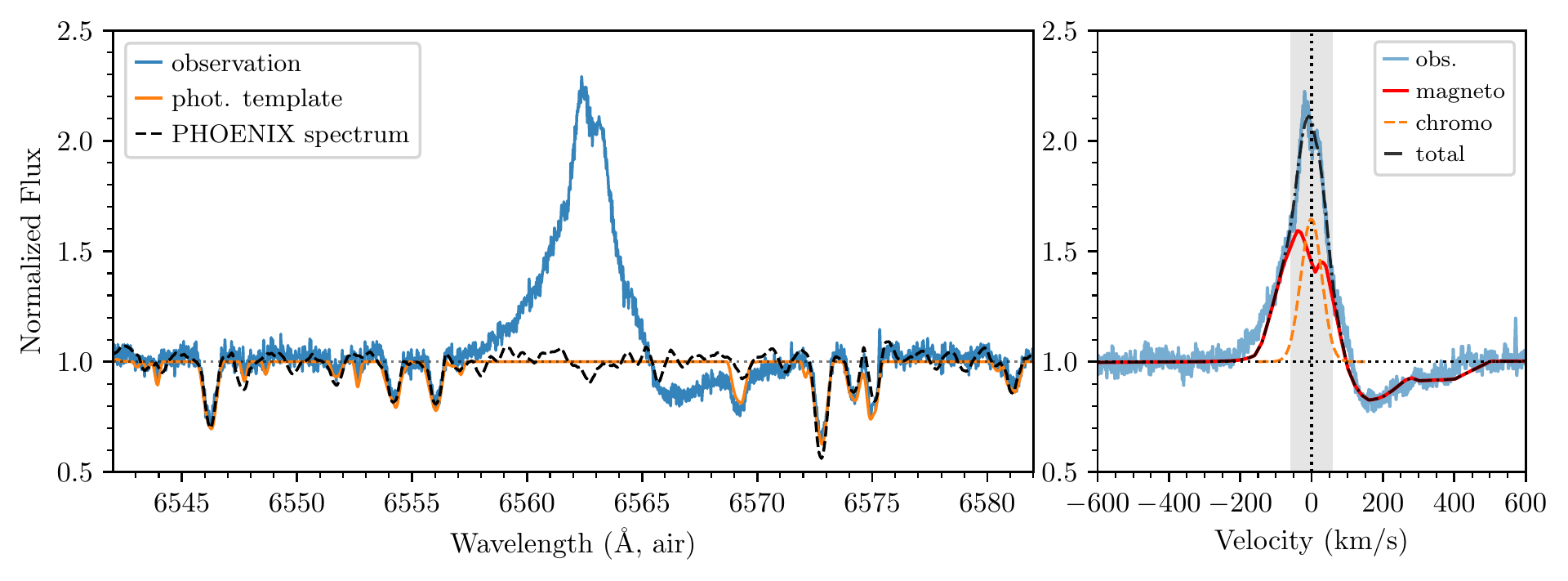}
\caption{A representative {\halpha} line profile of PDS~70. Left: The line before subtraction of the photosphere. An adopted photospheric template, constructed from the star's spectrum during quiescence, is shown in orange. The emission feature is excluded in the construction of the photospheric template. In comparison, the black line shows a K7 template from the PHOENIX model. Right: The photospheric-subtracted line profile (blue) and the best fit model (red). The line center ($\pm\,60\,\kms$, shaded) is excluded from the magnetospheric model fit. The dashed orange line shows the best fit for the chromospheric profile, and the dash-dotted line shows the total model line profile. For this observation, the best fit parameters are ${\rm R_i=4.0\,R_{\star}}$, ${\rm W_r=0.6\,R_{\star}}$, \mdot=$1.0\times10^{-10}\,\msunyr$, \tmax=11,500\,K, and $i=45^{\circ}$. 
\label{fig:best_fit}}
\end{figure*}

The chromospheric contribution to
the hydrogen lines becomes significant in low accretors.
In addition, photospheric absorption lines can affect
the shape of the redshifted absorption features in the line. Therefore, the photospheric and chromospheric contributions to the line need to be taken into account before modeling the line profile.

We constructed a photospheric template of the star using its normalized spectrum at the most quiescent state, during which the {\halpha} line is symmetric and purely in emission. We replaced the {\halpha} emission feature and any small features within 30\% of the standard deviation of the flux  with ${\rm F_{\lambda} = F_{\lambda, norm} = 1}$. We then used a box filter to smooth the spectrum, resulting in a photospheric template.
The left panel of Figure~\ref{fig:best_fit} shows a spectrum of the star and the photospheric template derived from the stellar spectra. In comparison, we plot a photospheric template interpolated from the PHOENIX model spectra \citep{husser2013}, in the same spectral resolution and rotational velocity. The template is similar to the PHOENIX spectra, and it better reproduces the \ion{Fe}{1} absorption line at {6569.2\,\AA}.

We fit each observation by calculating the root-mean-square error (RMSE) for all profiles in the grid of models. The RMSE is given by 
\begin{equation} 
    \text{RMSE} = \sqrt{\frac{\sum_{i=1}^N\, \left(F_{{\rm obs}, i} - F_{{\rm model}, i}\right)^2}{N}}, 
\end{equation} 
where $F_i$s are normalized fluxes at any given pixel and $N$ is the total numbers of pixels in the relevant velocity range. 
This statistic avoids giving weight to any particular feature of the observed profile, unlike the $\chi^2$ statistics, which is biased toward emission features if the deviation is normalized by the observed flux at a given pixel. We only consider the velocity range of $-250$ to $+400\,\kms$, comparable to the star's free-fall velocity, and exclude the region $\pm60\,\kms$ to avoid fitting the chromospheric feature of the line. The best fit models are the models with the smallest RMSE, and the mass accretion rate and the accretion geometry is inferred from the weighted mean parameters of the model with RMSE $\leq$ 0.1. To verify that the line center is Gaussian, indicating that it arises in the chromosphere, we fit a Gaussian profile to the residual of the best fit, and then add the Gaussian profile to the photospheric profile. The right panel of Figure~\ref{fig:best_fit} shows an example of the best fit for one of the observed profiles. The observed line profile is reproduced by adding the best fit magnetospheric profile with a Gaussian profile. 
\footnote{Since determining the properties of the chromosphere is beyond the scope of this paper, we do not attempt to fit the line center with a double-Gaussian model typically employed to fit chromospheric lines \citep[e.g.][]{rauscher2006}.} The mass accretion rate of the star based on this observation is $1.0\times10^{-10}\,\msunyr$.

\subsubsection{Accretion Variability}
\begin{figure*}[t]
\epsscale{1.15}
\plotone{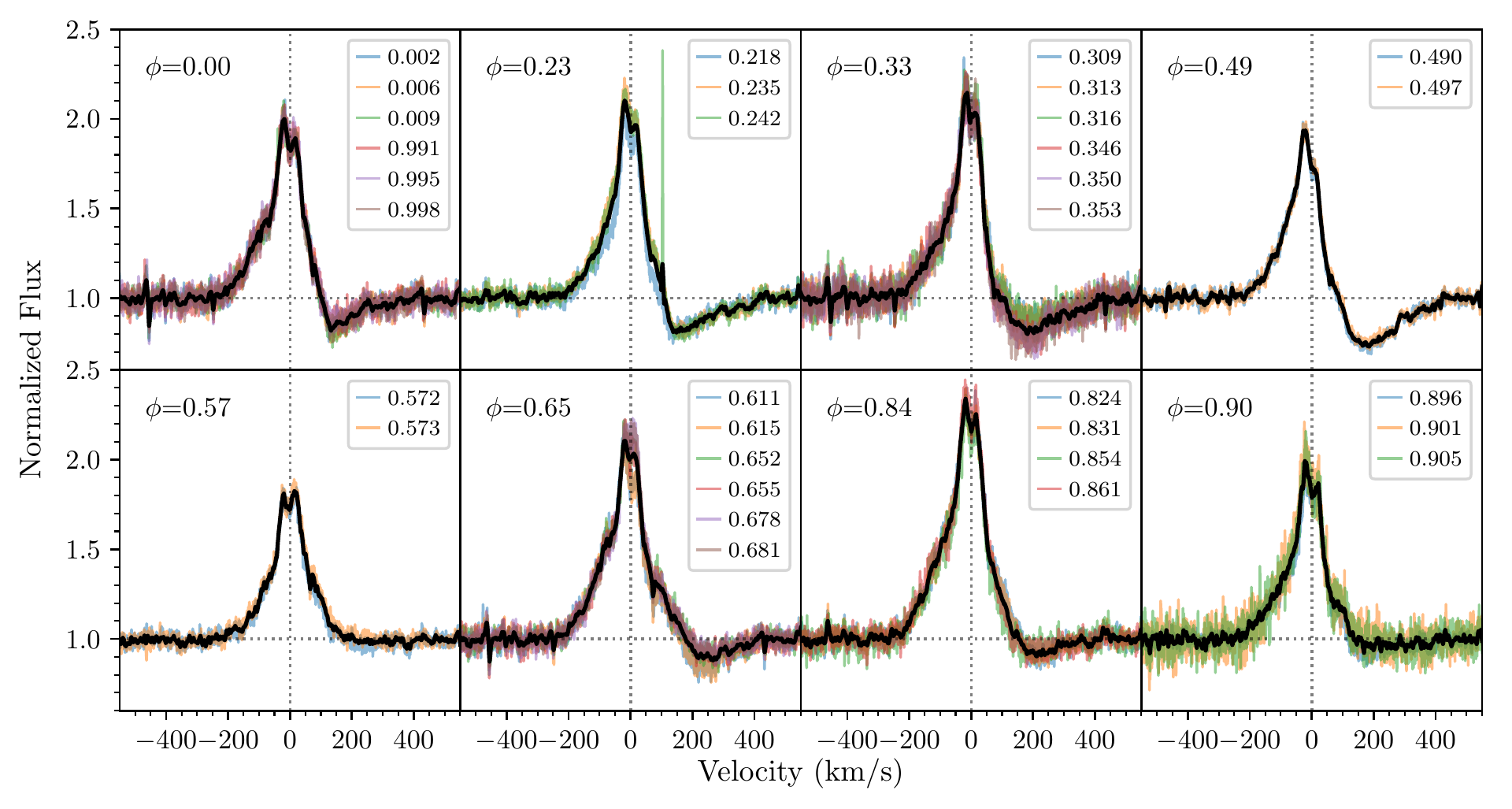}
\caption{The {\halpha} line profiles of PDS~70 grouped and stacked in 8 phases in the rotation period. Most of the spectra in the same group are from the same night, but spectra observed $\sim1$ rotation period apart shows remarkable similarity. The stacked spectra are smoothed with the Savitzky-Golay filter 21 pixels in size for clarity.
\label{fig:phases}}
\end{figure*}

To explore the variability of the mass accretion rate and accretion geometry, we grouped the 32 
HARPS spectroscopic observations by rotation phases,
calculated from the period in Section~\ref{sssec:rotation_period}, and stacked them, resulting in 8 spectra representing observations in different phases. The phase $\phi$ is defined such that $\phi=0$ at Modified Julian Date (MJD) 0.0.  The grouping was done such that observations on the same night fell on one group, and the profiles 
were similar. As shown in Figure~\ref{fig:phases}, the star shows rotational variability in the line profiles. From phase 0.00, the redshifted absorption component becomes stronger and is strongest at phase 0.49. On the other hand, the magnetospheric component of the line from phase 0.57 to 0.90 is mostly in emission, with weak to no redshifted absorption. 

To quantify the phase-variation of properties of the accretion flow, we fitted the stacked spectra at each phase with the grid of models, as described in \ref{sssec:fit_halpha}. The results are shown in Table~\ref{tab:model_results}, in which the last column shows the number of models within the RMSE$\leq0.1$ criterion. The last row of the Table shows the phase-averaged values for each model parameters calculated by weighting each observation with a phase duration $\Delta\phi$, where 
\begin{equation} 
    \Delta\phi_i = \frac{\phi_{i+1} - \phi_i}{2} + \frac{\phi_i - \phi_{i-1}}{2}. 
\end{equation} 
Except for the flow temperature {\tmax}, the model parameters show various degrees of variability. In particular, the observed {\mdot} can vary by more than a factor of three during the rotation phase.

\begin{figure*}[t]
\epsscale{1.15}
\plotone{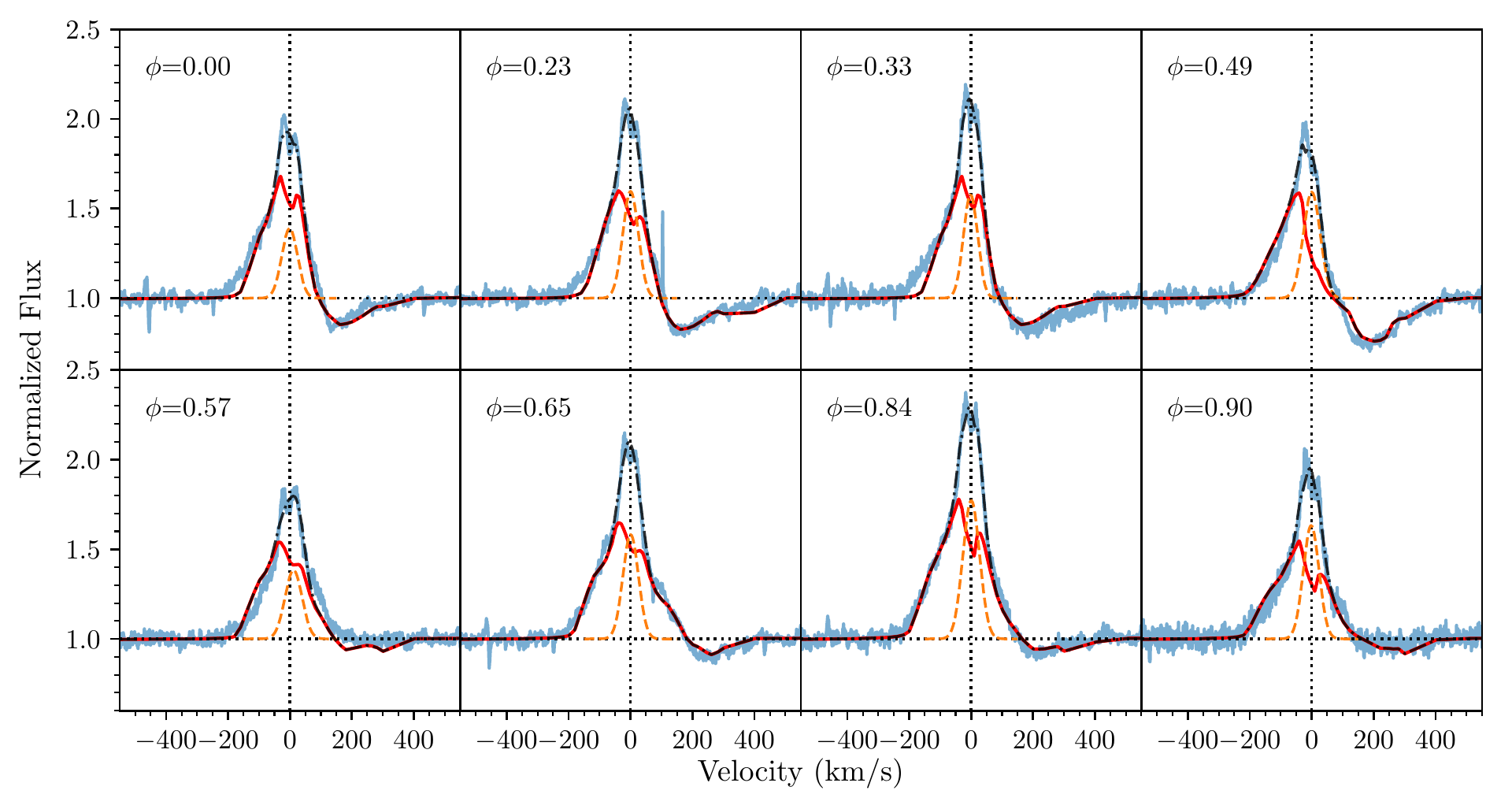}
\caption{Line profiles of the representative best fits for each of the observed phases. The observation is shown in blue, and the magnetospheric model is shown in red. The dashed orange lines are the chromospheric contributions, with the total model profile shown in dashed-dot black lines.
\label{fig:best_fits_phase}}
\end{figure*}

Figure~\ref{fig:best_fits_phase} shows a representative best fit for each of the observed profiles. As shown in the figure, our best fit models show good agreement with the observed line profiles. The chromospheric fits to the residual (dashed orange lines) are centered at $v\sim0\,\kms$, as expected, except at $\phi=0.57$. In this case, the magnetospheric model could not well reproduce the symmetric feature of the observed profile and tends to under-predict the emission on the red wing, causing the residual (from which the chromosphere is fitted) to be redshifted. Since redshifted absorption suggests material flowing along the line of sight, the symmetric line profile suggests that there is a region of the magnetosphere with little to no flow, which is passing in front of the star at the given phase. The modifications to the current model required to test this hypothesis are beyond the scope of this paper.

\begin{figure}[t]
\epsscale{1.15}
\plotone{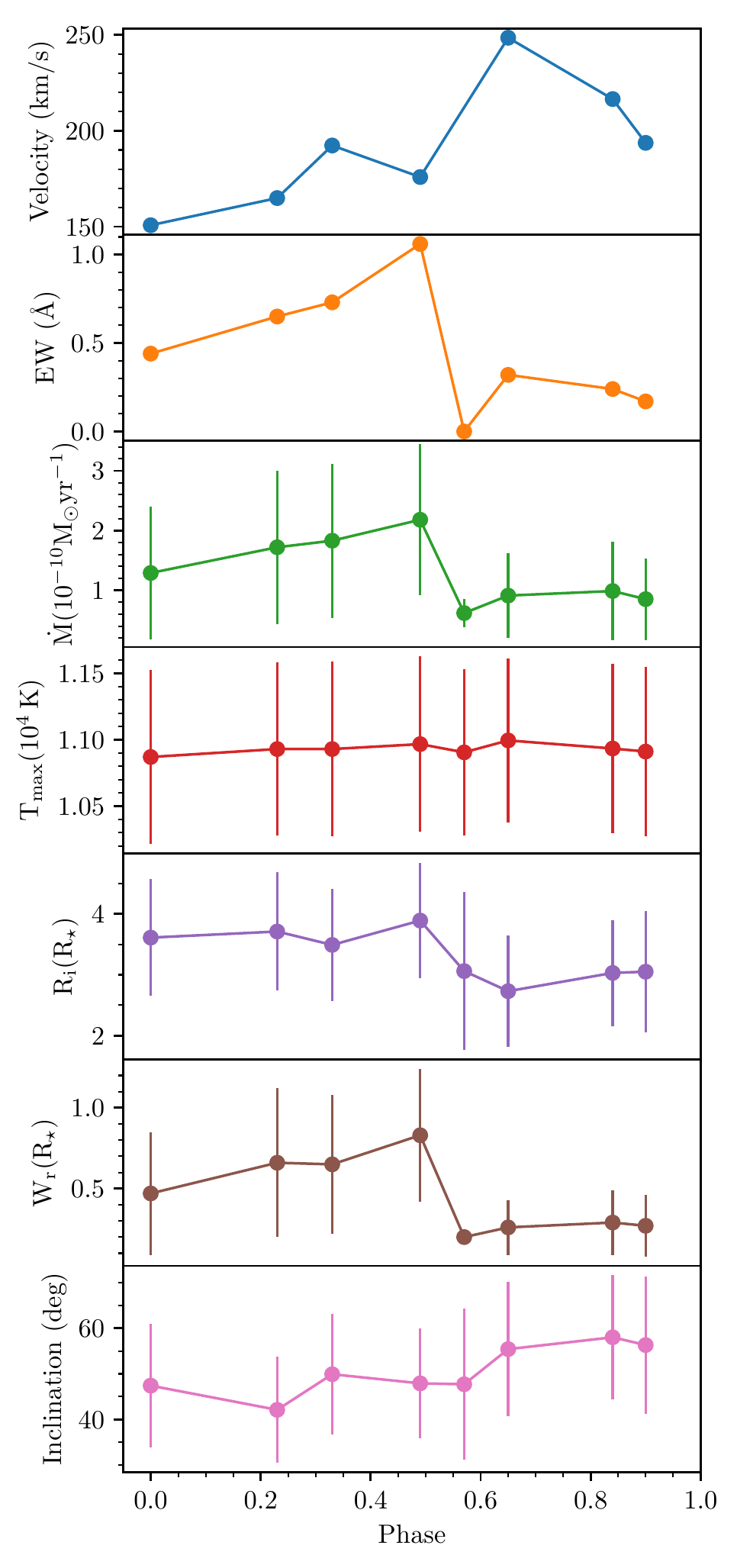}
\caption{Phase variation of the properties of the redshifted absorption component in the observed line profile (top two rows) and that of the magnetospheric accretion model fit parameters. The velocity of the redshifted absorption is excluded for phase 0.57 since the feature is undetected.
\label{fig:phases_velocity}}
\end{figure}

We compare the best-fit parameter values and the properties of the observed line profiles as a function of phase in Figure~\ref{fig:phases_velocity}. The top two rows of the Figure show the velocity at the lowest flux and the equivalent width (EW) of the redshifted absorption feature of the {\halpha} line, respectively. While the flow temperature is constant, general trends can be seen for other model parameters. When the absorption is weak with the smallest EW, the mass accretion rate is low, the truncation radius (\ri) is small, and the magnetosphere is thin. Thicker and larger magnetospheres correspond to stronger absorption features and higher mass accretion rates. The model inclination $i$, which probe the inclination of the magnetic axis, seems to vary slightly with a similar trend seen in the redshifted velocity. The slight variation in the redshifted velocity, and to some extent the inclination, seems to suggest that there could be a small misalignment between the magnetic axis and the rotation axis, resulting in a non-axisymmetric accretion flow. Numerical simulation of accreting magnetized star with misaligned axis \citep{romanova2003} suggests that even a small misalignment ($\Delta i \sim 5^{\circ}$) results in non-axisymmetric flow, with mass preferentially flow along a particular path. We note, however, that the uncertainty in the inclination is large compared to the model variation. Therefore, this alone, without the variability of other observed and model parameters, would not suggest the misalignment.

The variation of the model parameters suggests that we may be probing different portions of this asymmetric flow. However, as our current model does not allow misalignment between the stellar rotation axis and the magnetic axis, we caution that such effects have to be further investigated in the future. Nevertheless, our model still provides an estimate of the mass accretion rate and its possible variation. Future spectropolarimetric observations would provide insight into the geometry of the magnetosphere \citep{donati2007,donati2011a}.

We also note that in Figure~\ref{fig:best_fits_phase}, the chromospheric components is varying slightly between phases, especially the strength of the line, while the width and the line center are constant. We defer a discussion about chromospheric component to future studies.

\begin{deluxetable*}{ccccccc}
\tablecaption{Results of the Magnetospheric Accretion Model \label{tab:model_results}}
\tabletypesize{\scriptsize}
\tablehead{
\colhead{Phase} &
\colhead{\mdot} &
\colhead{\ri} &
\colhead{\rw} &
\colhead{\tmax} &
\colhead{$i$} &
\colhead{Numbers} \\
\colhead{} &
\colhead{($10^{-10}\,\msunyr$)} &
\colhead{(R$_{\star}$)} &
\colhead{(R$_{\star}$)} &
\colhead{($10^4\,$K)} &
\colhead{(deg)} &
\colhead{of models}
}
\startdata
0.00  &  1.3$\pm$1.1  &  3.6$\pm$1.0  &  0.5$\pm$0.4  &  1.09$\pm$0.07  &  47$\pm$14  &  1229  \\
0.23  &  1.7$\pm$1.3  &  3.7$\pm$1.0  &  0.7$\pm$0.5  &  1.09$\pm$0.07  &  42$\pm$12  &  1548  \\
0.33  &  1.8$\pm$1.3  &  3.5$\pm$0.9  &  0.7$\pm$0.4  &  1.09$\pm$0.07  &  50$\pm$13  &  1441  \\
0.49  &  2.2$\pm$1.3  &  3.9$\pm$0.9  &  0.8$\pm$0.4  &  1.10$\pm$0.07  &  48$\pm$12  &  1333  \\
0.57  &  0.6$\pm$0.2  &  3.1$\pm$1.3  &  0.2$\pm$0.0  &  1.09$\pm$0.06  &  48$\pm$17  &  259  \\
0.65  &  0.9$\pm$0.7  &  2.7$\pm$0.9  &  0.3$\pm$0.2  &  1.10$\pm$0.06  &  55$\pm$15  &  432  \\
0.84  &  1.0$\pm$0.8  &  3.0$\pm$0.9  &  0.3$\pm$0.2  &  1.09$\pm$0.06  &  58$\pm$14  &  483  \\
0.90  &  0.8$\pm$0.7  &  3.0$\pm$1.0  &  0.3$\pm$0.2  &  1.09$\pm$0.06  &  56$\pm$15  &  571  \\
\hline
Average & 1.3$\pm$0.5 &  3.4$\pm$0.4  &  0.5$\pm$0.2  &  1.09$\pm$0.00  &  50$\pm$6  &  \\
\enddata
\end{deluxetable*}

\section{Discussion}

\subsection{Accretion shock emission}

Our variability analysis indicates that the mass accretion rate onto PDS~70 is in the range $0.6 - 2.2  \times 10^{-10}\msunyr$ (Table~\ref{tab:model_results}). Material accreting at this rate is slowed down through an accretion shock at the stellar surface before merging with stellar material. Emission from this shock shows more conspicuously in the UV \citep{calvet1998,hartmann2016}.

\begin{figure}[t!]
\epsscale{1.15}
\vspace{0.05in}
\plotone{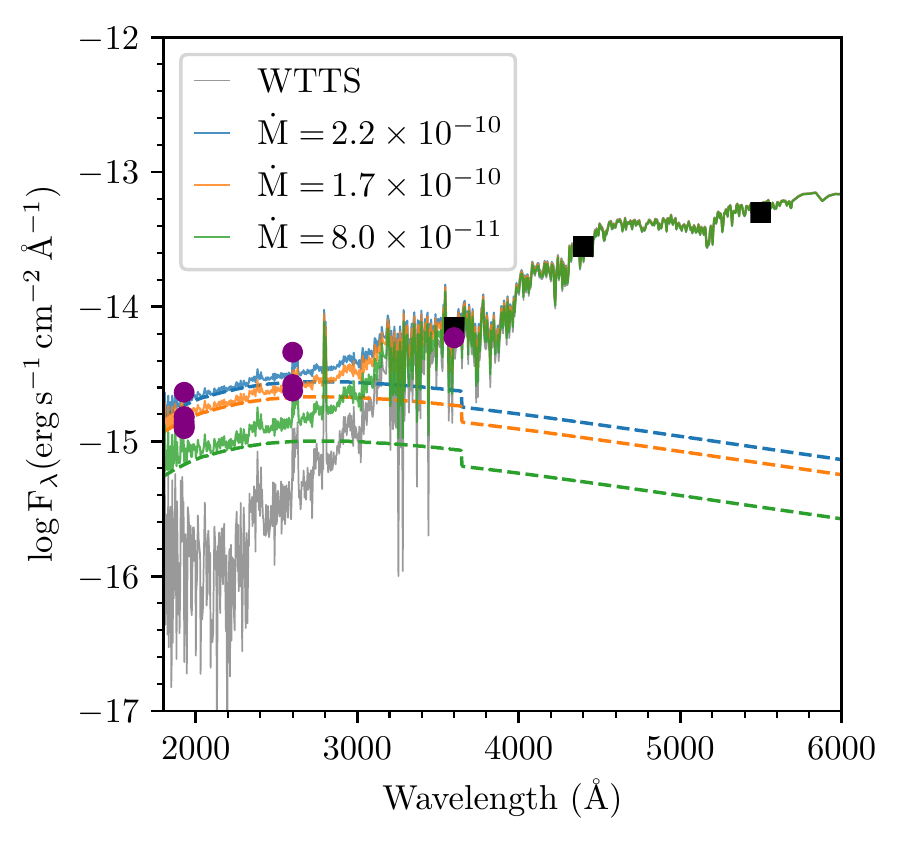}
\caption{
Optical and UV fluxes of PDS~70 and model spectra, including emission from the accretion shock. The optical fluxes (black squares) are taken from \citet{gregorio-hetem2002} and the Swift U, uvw1, and uvw2 fluxes (purple) from \citet{joyce2019}. The gray line is the spectrum of the weak T Tauri star HBC 427 with the same spectral type as that of PDS~70; we added the expected emission from the accretion shock to this spectrum. The solid lines show the total emission for different mass accretion rates and the dashed line of the same color showing only the shock emission.
\label{fig:shock}}
\end{figure}

We show in Figure \ref{fig:shock} optical fluxes for PDS~70 from \citet{gregorio-hetem2002} and fluxes in the Swift/UVOT U, uvw1, and uvw2 bands \citep{joyce2019}. Except for one case, observations at uvw1 and uvw2 were obtained at different epochs. The V-I colors are similar to those of a K7 star from \citet{pecaut2013}, indicating no extinction.

We also show in Figure \ref{fig:shock} the total flux expected for PDS~70 when the emission from accretion shocks with $\Mdot$ between $0.8 \times 10^{-10}\,\msunyr$ and $2.2 \times 10^{-10}\,\msunyr$ are added to the stellar flux. The accretion shock emission is calculated using the methods of \citet{calvet1998} for the PSD~70 mass and radius, and for one characteristic value of the energy flux in the accretion column $ \curf = 10^{12}\, {\rm erg \, cm^{-2}\, s^{-1}}$ \citep{ingleby2013}. The intrinsic stellar spectrum is taken from the Weak T Tauri star HBC 427 \citep{ingleby2013}, scaled to the stellar radius and distance. The WTTS spectrum includes emission from the stellar chromosphere, which is strong in young stars \citep{ingleby2011a}. As shown in Figure \ref{fig:shock}, the observed Swift fluxes are consistent with the accretion rates estimated from the magnetospheric modeling of {\halpha} (Table \ref{tab:model_results}).

\citet{joyce2019} obtained an accretion rate
$\Mdot \sim 6\times10^{-12}\,\msunyr$ using the correlation of the flux
excess in the U band  and the accretion luminosity from \citet{venuti2014}. As can be
seen in Figure \ref{fig:shock}, the flux at the U band is dominated by stellar photospheric and chromospheric emission,
as expected for low accretors \citet{ingleby2011a}, so we argue that this band is not optimal
to obtain accretion shock emission in this star. 
In general, relationships between the U excess and the accretion luminosity cannot be calibrated at low levels of accretion and should not be used.

\subsection{The stellar mass accretion rate as a potential diagnostic of disk accretion processes}
Our analysis suggests that PDS~70 is accreting at a moderate rate of $\sim10^{-10}~M_\odot{\rm yr}^{-1}$. Where does this mass come from? Does the outer disk gas flow across the cavity, or is there a mass reservoir to sustain the stellar accretion in the inner disk?

Numerical simulations have shown that the outer disk gas, beyond the gap opened by planets, can flow into the inner regions of the disk. The mass flow rate depends upon various factors, including the number of planets responsible for the gap, planets' masses, planets' accretion efficiency, and protoplanetary disk thermodynamics \citep{lubow2006,zhu2011,muller2013}. In general, these numerical studies show that the mass flow rate across the gap is $1-100~\%$ of the mass accretion rate beyond the gap \citep{lubow2006,zhu2011,muller2013}.

\begin{figure}[t!]
\epsscale{1.15}
\vspace{0.05in}
\plotone{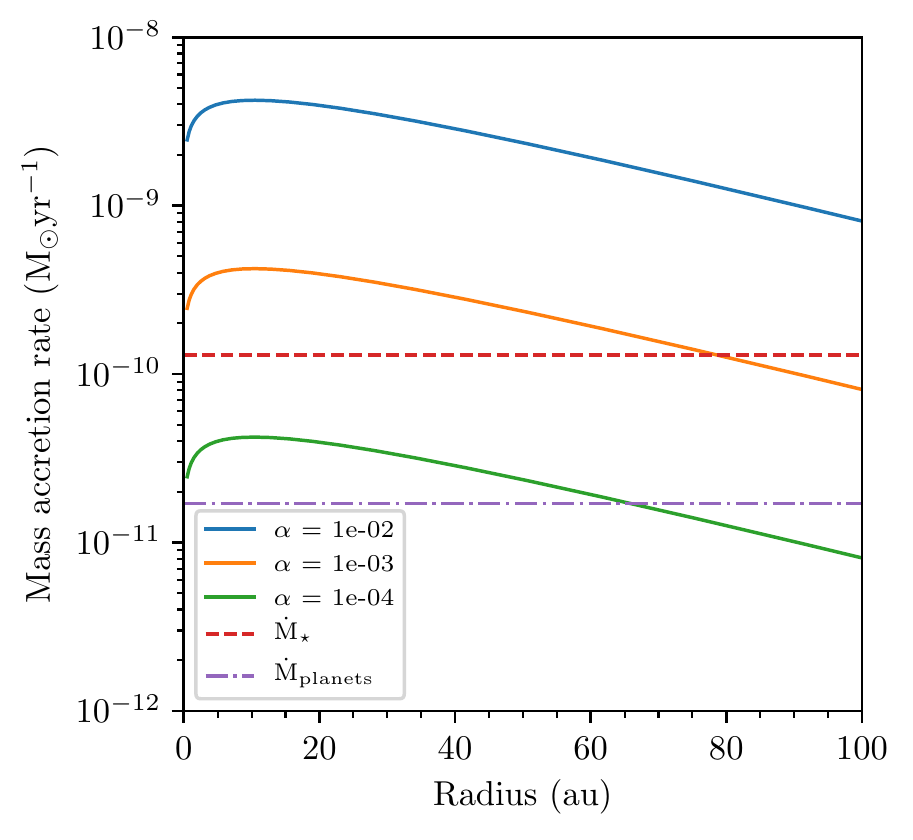}
\caption{
The accretion rate of a viscous disk, adopting disk surface density and temperature profiles described in Equations (\ref{eqn:disk_density}) and (\ref{eqn:disk_temperature}). Three different viscosity parameter values are assumed: $\alpha=10^{-2}, 10^{-3}$, and $10^{-4}$. The horizontal lines show the mass accretion rate of the star (red dashed line) and the total mass accretion rates of the planets (purple dash-dotted line).
\label{fig:mdot_disk}}
\end{figure}

In order to examine if the outer disk gas can explain PDS~70's accretion rate, we first calculate the accretion rate of an unperturbed viscous accretion disk. We adopt the disk surface density and temperature profiles used in \citet{bae2019}:
\begin{equation}
\label{eqn:disk_density}
    \Sigma(R) = 2.7~{\rm g~cm}^{-2} \left(\frac{R}{40~{\rm au}}\right)^{-1} \exp\left(- {R \over 40~{\rm au}} \right)
\end{equation}
and
\begin{equation}
\label{eqn:disk_temperature}
    T(R) = 38~{\rm K} \left( {R \over 40~{\rm au}} \right)^{-0.24}.
\end{equation}
In Figure \ref{fig:mdot_disk}, we present the disk accretion rate, calculated as $\dot{M}_d = 3 \pi \nu \Sigma$ where $\nu$ is the disk kinematic viscosity defined as $\nu = \alpha c_s^2/\Omega$. Here, $\alpha$ is the Shakura-Sunyaev viscosity parameter characterizing the mass transport efficiency \citep{shakura1973}, $c_s$ is the disk sound speed, and $\Omega$ is the angular frequency. At 70~au, beyond the common gap opened by PDS~70b and c, the accretion rate is
\begin{equation}
  \dot{M}_d =  1.5\times10^{-10} M_\odot{\rm yr}^{-1} \left( {\alpha \over 10^{-3}} \right).
\end{equation}

If the disk accretion is efficient (i.e., $\alpha \gtrsim 10^{-3}$) so the outer disk supplies gas at $\gtrsim10^{-10}~M_\odot~{\rm yr}^{-1}$, it is thus possible that the stellar accretion is sustained by the mass reservoir in the outer disk, as the two planets within the gap are known to take only a small fraction of the supply ($\sim10^{-11}\,\msunyr$; \citealt{wagner2018,haffert2019,thanathibodee2019a,aoyama2019}). However, when the mass flow rate across the gap is significantly reduced and/or the disk has a low mass transport efficiency (i.e., $\alpha \ll 10^{-3}$), the stellar accretion rate of $\sim10^{-10}\,\msunyr$ is difficult to explain with the outer disk mass reservoir. In this case, we may need to invoke an inner disk mass reservoir (e.g., dead zone) that can feed the star for a prolonged period of time with a low efficiency \citep{hartmann2018}. The presence of compact sub-millimeter continuum emission shown in ALMA observations \citep{long2018a,keppler2019} may support this inner reservoir scenario.

\citet{manara2019} used the
planet population synthesis models of \citet{mordasini2009,mordasini2012b} to make predictions for stellar mass accretion rates and disk masses and compare them with observations of the Lupus and Chamaeleon star-forming regions. Their disk models assume a viscosity parameter of $\alpha=2\times10^{-3}$, and planets accrete the disk gas at a fraction of the disk viscous accretion rate. With the assumed viscosity parameter, the stellar accretion could be sustained by the outer disk reservoir, subject to a decrease in the presence of planets.

Their models, however, produce a larger fraction of weak accretors than observed in transition disks. As they already pointed out, one possible explanation to this conflict is that their prescription of gas accretion onto planets over-predicts the real accretion rate, and thus reduces the mass flow rate across the gaps more than it actually would. Instead, as we suggested above, this could be reconciled with an inner disk reservoir with a small $\alpha$, as the stellar accretion in this case would be less sensitive to the formation of giant planets in the outer disk. It will be interesting to run low-viscosity counterparts of planet population synthesis models to explore this possibility.

In summary, since our current understanding of protoplanetary disk accretion physics is incomplete, we cannot conclude whether PDS~70's accretion rate is sustained by an inner or an outer disk reservoir. We note that our calculations here are based on an assumption that the disk has a uniform $\alpha$;
other possibilities will be explored in future work. Future observations that can characterize the inner disk properties and search for potential inner disk winds, together with those that can constrain the level of turbulence in the outer disk, will help better understand the origin of PDS~70 accretion. In addition, observational searches of low accretors and accurate determination of $\Mdot$ \citep{thanathibodee2018,thanathibodee2019b} will help find the observational low limit of {\mdot} to compare with expectations of exoplanets population models.

\subsection{Origin of the Blueshifted Absorption in {\henir}}
The presence of a sub-continuum blueshifted absorption feature in the {\henir} line has been attributed to winds \citep{edwards2003,edwards2006}. In particular, the narrow blueshifted absorption is interpreted as a wind coming from the inner disk. The co-existence of blueshifted and redshifted absorption with very weak emission in the line center has been observed for a few stars in the high-resolution survey of {\henir} line \citep{edwards2006}. However, most of the stars in the surveys, and all of the stars with blue+red absorption, have high levels of accretion, {$\Mdot\geq10^{-8}\,\msunyr$}. Interestingly, PDS~70, with its very low accretion level, also shows a similar type of profile. In fact, in our ongoing survey of very low accretors, more than $\sim$10\% of disk-bearing K-M TTS with weak {\halpha} show blue+redshifted absorption. The velocity of the blueshifted absorption ($\sim-85\,\kms$) is consistent with the wind coming from the inner disk. 

We can estimate the mass loss rate from the wind by adopting the procedure outlined in \citet{calvet1997}. For a line formed in the wind, the optical depth at a given velocity $v$ is given by 
\begin{equation} 
    \tau = \frac{\pi e^2}{m_ec}\frac{f c}{\nu_0}\frac{n_l(v)}{dv/dz}, 
\end{equation} 
where $f$ is the oscillator strength, $\nu_0$ is the line frequency, $n_l$ is the population of the lower level, and $dv/dz$ is the velocity gradient. The mass loss rate is given by 
\begin{equation} 
    \Mdot_w \sim \Delta A v \mu m_H n_H(v) \sim \Delta A v \mu m_H \eta n_l(v), 
\end{equation} 
where $\Delta A$ is the cross-section area of the wind at $v$, $\mu$ is the mean molecular weight, $n_H$ is the number density of hydrogen, and $\eta \equiv n_H/n_l$. To the first approximation, $dv/dz \sim v/R_{\star}$ and $\Delta A \sim \pi (2R_{\star})^2$ \citep{calvet1997}. We estimate the parameter $\eta$ by calculating a typical fraction between the number density of the lower level of {\henir} to the total hydrogen number density. We use the C17.01 release of the software Cloudy \citep{ferland2017} to calculate the level populations in a slab of gas assuming that the ionization radiation is an X-ray with a blackbody temperature of $5\times10^6$\,K and $L_X = 5\times10^{29}\,{\rm erg \, s^{-1}}$ consistent with the luminosity measured by Swift \citep{joyce2019}.\footnote{\citet{joyce2019} model the X-ray spectra with two temperatures. Here we adopted a temperature inside the range of those temperatures.} 
Assuming that the wind is $\sim 10\,R_{\star}$ from the star with a thickness of $1\,R_{\star}$, we find that $\eta\sim10^7$ for a relevant range of $n_H$ across the slab. With $\tau\sim0.5$, estimated from the depth of the feature, and $\mu=2.4$, we find that $n_l(v=85\,\kms)\sim55\,{\rm cm^{-3}}$ and $\Mdot_w\sim10^{-11}\,\msunyr.$ This mass loss rate is consistent with that expected from a MHD inner disk wind in which $\Mdot_w \sim 0.1 \Mdot_{acc}$, suggesting that the blueshifted absorption in {\henir} forms by a similar mechanism as in high accretors \citep{calvet1997,edwards2006,kwan2007}. Since the blueshifted velocity is high, it is unlikely that the feature is formed in photoevaporative winds \citep[c.f.][]{alexander2014}. The features in the {\henir} lines of low accretors will be described in detail in a future study (Thanathibodee et al. in prep.)

Our detection of wind and accretion
signatures
confirms the existence of 
gas in the inner disk around PDS~70.
\citet{long2018a} searched for
first overtone CO lines in low resolution
near-IR spectra and could not
find them, from which they inferred that the
inner disk of PDS 70 was gas poor. However, the lack of detection could be due to the difficulty of separating the disk emission from the intrinsic photospheric CO absorption lines
\citep{calvet1991a}. Observations of ground state CO lines or of fluorescent H$_2$ lines can
help confirm the presence of gas in the inner disk of PDS70.


\section{Summary and Conclusions}

We have analyzed TESS photometry, archival HARPS spectra, and a FIRE near-IR spectrum of PDS~70, a $\sim$ 5\,Myr star with two confirmed giant planets forming in its circumstellar disk. The TESS variability is consistent with rotational modulation of spots on the stellar surface, indicating magnetic activity as found in other young stars. The period derived from TESS observations and the measured $v\sin i$ yield an inclination to the line of sight consistent with the disk inclination derived from submillimeter data, indicating that the rotational axes of the star and the disk are parallel to each other within uncertainties. We find redshifted absorption features in the {\henir} line and in {\halpha}, confirming that the star is accreting. We model the {\halpha} profiles assuming magnetospheric accretion and find mass accretion rates in the range $0.6 - 2.2 \times 10^{-10}\,\msunyr$. These values of {\mdot} predict a UV flux from the surface accretion shocks consistent with the flux observed in Swift UV bands. We analyze changes in the geometry of the magnetospheric flows with rotation phase and find that it could be non-axisymmetric, consistent with a small tilt between the stellar rotation axis and the magnetic axis. The relatively high values of the mass accretion rate may indicate the need for an additional mass reservoir in the disk to feed the flows onto the star. We estimate the mass loss rate from the blueshifted absorption feature in the {\henir} line and find that the rate and the velocity of the line are consistent with the wind being driven by accretion. The detection of accretion and winds confirms the existence of 
gas in the inner disk.

\acknowledgments
\noindent
We thank A. Katherina Vivas for enlightening us on period determinations, and Charles Cowley for insightful discussions on the {\henir} line. This project is supported in part by NASA grant NNX17AE57G. JH acknowledges support from program UNAM-DGAPA-PAPIIT grant IA102319. 

This paper is based on data obtained from the ESO Science Archive Facility under request number thanathi/498913. This paper includes data collected with the TESS mission, obtained from the MAST data archive at the Space Telescope Science Institute (STScI). Funding for the TESS mission is provided by the NASA Explorer Program. STScI is operated by the Association of Universities for Research in Astronomy, Inc., under NASA contract NAS 5–26555. 
This research made use of Astropy,\footnote{http://www.astropy.org} a community-developed core Python package for Astronomy \citep{astropy2013,astropy2018}.

\facilities{TESS, Magellan:Baade (FIRE), ESO:3.6m (HARPS)}

\software{Lightkurve \citep{lightkurve-collaboration2018}, Psearch code \citep{saha2017}, FIRE data reduction pipeline \citep{simcoe2013}, Astropy \citep{astropy2013,astropy2018}}

\newpage
\bibliography{tt_rev1}{}
\bibliographystyle{aasjournal}

\end{document}